\documentclass[english,12pt,draftcls, onecolumn]{IEEEtran}

\usepackage[T1]{fontenc}
\usepackage[latin9]{inputenc}
\usepackage{amsmath}
\usepackage{graphicx}
\usepackage{amssymb}
\makeatletter

\providecommand{\tabularnewline}{\\}
\newcommand{\lyxdot}{.}

\def\d{\mathrm{d}}

\def\r{\eta}

\newtheorem{definitn}{Definition}
\newtheorem{prop}{Proposition}
\newtheorem{lemma}{Lemma}

\usepackage{babel}
\makeatother

\begin{document}

\title{Dynamic Connectivity in ALOHA Ad Hoc Networks}
\author{\begin{tabular}[t]{c@{\extracolsep{8em}}c}
  Radha Krishna Ganti  & Martin Haenggi\\
  Dept. of Electrical Engineering & Dept. of Electrical Engineering\\
University of Texas at Austin & University of Notre Dame\\
 Austin, TX 78712-0204, USA & Notre Dame, IN 46556, USA\\
 rganti@austin.utexas.edu &  mhaenggi@nd.edu
\end{tabular}}


\maketitle
\begin{abstract}
In a wireless network  the set of transmitting nodes  changes frequently
because of the  MAC scheduler and the traffic load. Previously, connectivity
in wireless networks was analyzed using static geometric graphs, and as we show
leads to an overly constrained design criterion. The dynamic nature of the
transmitting set introduces additional randomness in a wireless system that
improves the connectivity, and this additional randomness is not
captured by a static connectivity graph. In this paper,
we consider an ad hoc network with half-duplex radios that uses multihop routing and slotted
ALOHA for the MAC contention and introduce a random dynamic
multi-digraph to model
its  connectivity. We first provide analytical results about the degree distribution of
the graph. Next, defining  the  path formation time  as the minimum time required
for a causal path to form between the source and destination on the  dynamic graph, we derive
the distributional properties of the connection delay  using techniques from
first-passage percolation and epidemic processes. We consider the giant
component of the network formed when communication is noise-limited (by
neglecting interference). Then, in the presence of interference, we prove that the
delay scales linearly with the source-destination distance on this giant
component.  We also provide
simulation results to support the theoretical results.
\end{abstract}

\section{Introduction}
In a multihop ad hoc network, bits, frames or packets are transferred
from a source to a destination in a multihop fashion with the help
of intermediate nodes. Decoding, storing, and relaying introduces a delay that, measured in time slots, generally exceeds the number of hops.  For example, a five-hop route does not guarantee a delay of only five time slots. In a general setting, each
node can connect to multiple nodes. So a large number of paths may
form between the source and the destination. Each path may have taken
a different time to form with the help of different intermediate nodes.
Consider a network in which each node wants to transmit to its destination
in a multihop fashion. In general in such a network, a relay node
queues the packets from other nodes and its own packets and transmits
them according to some scheduling algorithm. If one introduces the
concept of queues, the analysis of the system becomes extremely complicated
because of the intricate spatial and temporal dependencies between
various nodes. In this paper we take a different approach. We are
concerned only with  the physical connections between nodes, i.e.,
 we do not care when a node $i$ transmits a particular packet
to a node $j$ (which depends on the scheduler), but we analyze when
a (physical) connection (maybe over multiple hops) is formed between
the nodes $i$ and $j$. This delay is a lower bound on the delay
with any queueing scheduler in place.

We assume that the nodes are distributed as a Poisson point process
(PPP) on the plane. In each time slot, every node decides to transmit
or receive using ALOHA. Any transmitting node can connect to a receiving
node when a modified version of the  protocol model criterion introduced in \cite{gupta2000cwn} is met.
Since at each time instant, the transmitting and receiving  nodes change,
the  connectivity graph changes dynamically. We analyze the time required
for a causal path to form between a source and a destination node.
The system model is made precise in Section \ref{sec:System-Model}.

This problem is similar in flavor to the problem of First-Passage
Percolation (FPP) \cite{kesten1986afp,hammersley1965fpp,aldous2003pds},
and the process of dynamic connectivity also resembles an epidemic
process \cite{durrett-1999,mollison1977scm,mollison1978mcp} on a
Euclidean domain. In a spatial epidemic process, an infected individual
infects a certain (maybe random) neighboring population, and this
process continues until the complete population is infected or the
spreading of the disease stops. In  the literature cited above,   the spreading
time of
the epidemic   is analyzed for different models of disease spread. We draw many ideas from this theory of epidemic
process and FPP. The main difference between an epidemic process and
the process we consider is that the spreading (of packets) depends on a
subset of the population (due to interference) and is not independent
from node to node.
In \cite{dousse2004lws}, the
latency for a message to propagate in a sensor network is analyzed using similar
tools. They consider a Boolean connectivity model with randomly weighted
edges and derive the properties of first-passage paths on the weighted
graph. Their model does not consider interference and thus allows
the use of Kingman's subadditive ergodic theorem \cite{kingman1973set}
while ours does not.  Percolation in  signal-to-interference ratio graphs was analyzed in \cite{dousse2006psi} where the nodes are assumed to be full-duplex.
In practice, radios do not transmit  and receive at the same time (at the same frequency), and hence the instantaneous
network  graph is always disconnected. In
\cite{ganti2009bounds, ganti2007dynamic}, we have introduced the concept of
dynamic connectivity graphs, and we proved that the average delay scales linearly
with source-destination distance  but the temporal correlation between
interference was neglected.  Baccelli et al. introduced a similar
concept of SINR-time graphs for ALOHA networks \cite{baccelli-stochastic}
wherein they proved that below a certain ALOHA parameter $p$, the average delay
of connectivity between nodes scales linearly with the distance by considering
the temporal correlation of the interference. In this paper we show a similar
result for the protocol model of communication. We also show that for a
positive fraction of nodes, the time of connectivity scales linearly with
the source-destination irrespective of the ALOHA parameter.
Connectivity between nodes far apart occurs
because of the dynamic nature of the MAC protocol. We first introduce  a dynamic  graph process to  model and  analyze connectivity and then
 derive the properties of this graph process for ALOHA.

In Section \ref{sec:System-Model}, we introduce the system model.
In Section \ref{sec:snapshot}, we study the connectivity properties
of  the random geometric graph formed at any time instant. In Section \ref{sec:time-evolution}, we
derive the properties of the delay and the average number of paths
between a source and destination and  show that the delay increases
linearly with increasing source-destination distance or, equivalently,
that the propagation speed is constant, i.e., the distance of the farthest
nodes to which the origin can connect   increases linearly with time.

\section{\label{sec:System-Model}System Model}

The location of the wireless nodes (transceivers)  is  assumed to  be  a  Poisson point process (PPP) $\phi$
of intensity $\lambda$ on the plane. We   assume that   time is slotted and  the MAC protocol used is slotted  ALOHA.
In every time slot each node    transmits with probability  $p$. Nodes are half-duplex, and   they   act as  receivers if they are not transmitting.
We use the protocol model \cite{gupta2000cwn}   to decide if the communication between a transmitter and a receiver is successful in a  given time slot:  A transmitting node located at $x$
can connect to a receiver located at $y$ if two conditions are met:
\begin{enumerate}
\item \textit{Interference:} The disk $B(y,\beta\|x-y\|),\beta >0$,
does not contain any other transmitting nodes.
\item \textit{Noise:} $\|x-y\|<\r$.
\end{enumerate}
   $B(x,r)$ denotes a disk of radius $r$ centered around $x$ and $B^c(x,r)=\mathbb{R}^2\setminus B(x,r)$.   $\beta$  is a system parameter and   captures  the resilience of the receiver against interference.
     The standard physical SINR model of communication can be related to the protocol model easily  when there is no fading. A detailed discussion about the protocol model can be found in \cite{kumar2006sla}. An interference-limited regime can  be modeled by dropping condition 2. In a similar fashion, a noise-limited scenario can be modeled by dropping condition 1.

   We shall use  $\mathbf{1}(x\rightarrow y,\Delta,\r)$  to  represent a random variable that is equal
to one if a transmitter at $x$ is able to connect to a receiver
$y$ when the transmitting set is $\Delta$, i.e., the interfering set is $\Delta\setminus \{x\}$. We will drop $\Delta$ if there is no ambiguity.
 At any time instant $k$, we denote the set of transmitters (decided by ALOHA) by $\phi_{t}(k)$ and the
set of receivers by $\phi_{r}(k)$. So we have $\phi_{t}(k)\cup\phi_{r}(k)=\phi$ and $\phi_{t}(k)\cap\phi_{r}(k)=\emptyset$,	 where $\emptyset$ denotes the empty set.

\begin{figure}
\begin{centering}
\includegraphics[width=4.2 in]{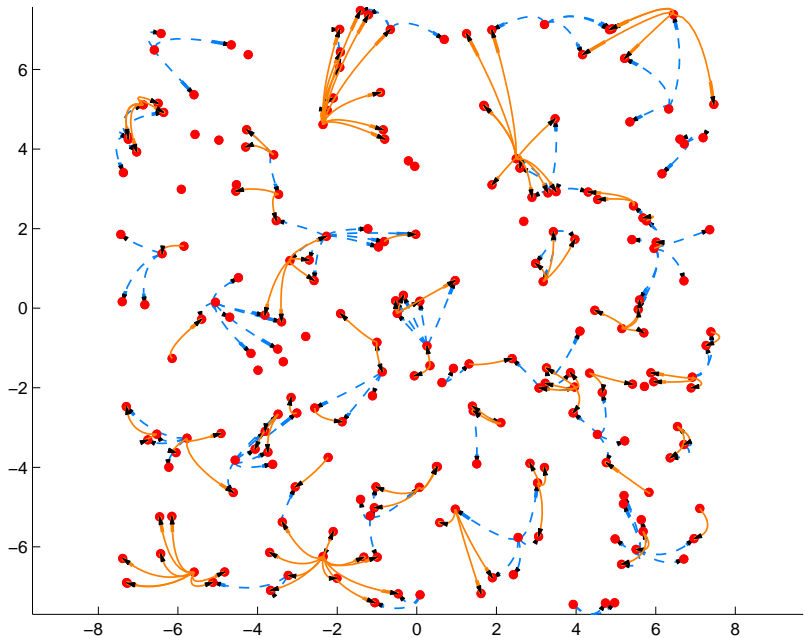}
\par\end{centering}
\caption{Illustration of $G(0,1)$, $p=0.2$, $\beta=1.2$. Dashed line represent edges in $g(0)$ (edges with  weight $0$) and solid lines represent  edges in $g(1)$ (edges with weight $1$).}
\label{fig:illus_dynamic}
\end{figure}

 The connectivity at time $k$ is captured  by a  directed  and weighted  random geometric graph $g(k)=\left(\phi,   E_{k}\right)$
with vertex set   $\phi$ and edge  set
\begin{equation}
 E_{k}=\left\{ (x,y)\colon \ \mathbf{1}\left(x\rightarrow y,\phi_{t}(k),\r\right)=1,x\in\phi_{t}(k),y\in\phi_{r}(k)\right\}.
\label{eq:def}
\end{equation}
\begin{figure}
\begin{centering}
\begin{tabular}{|c|c|}
\hline
\includegraphics[scale=0.4]{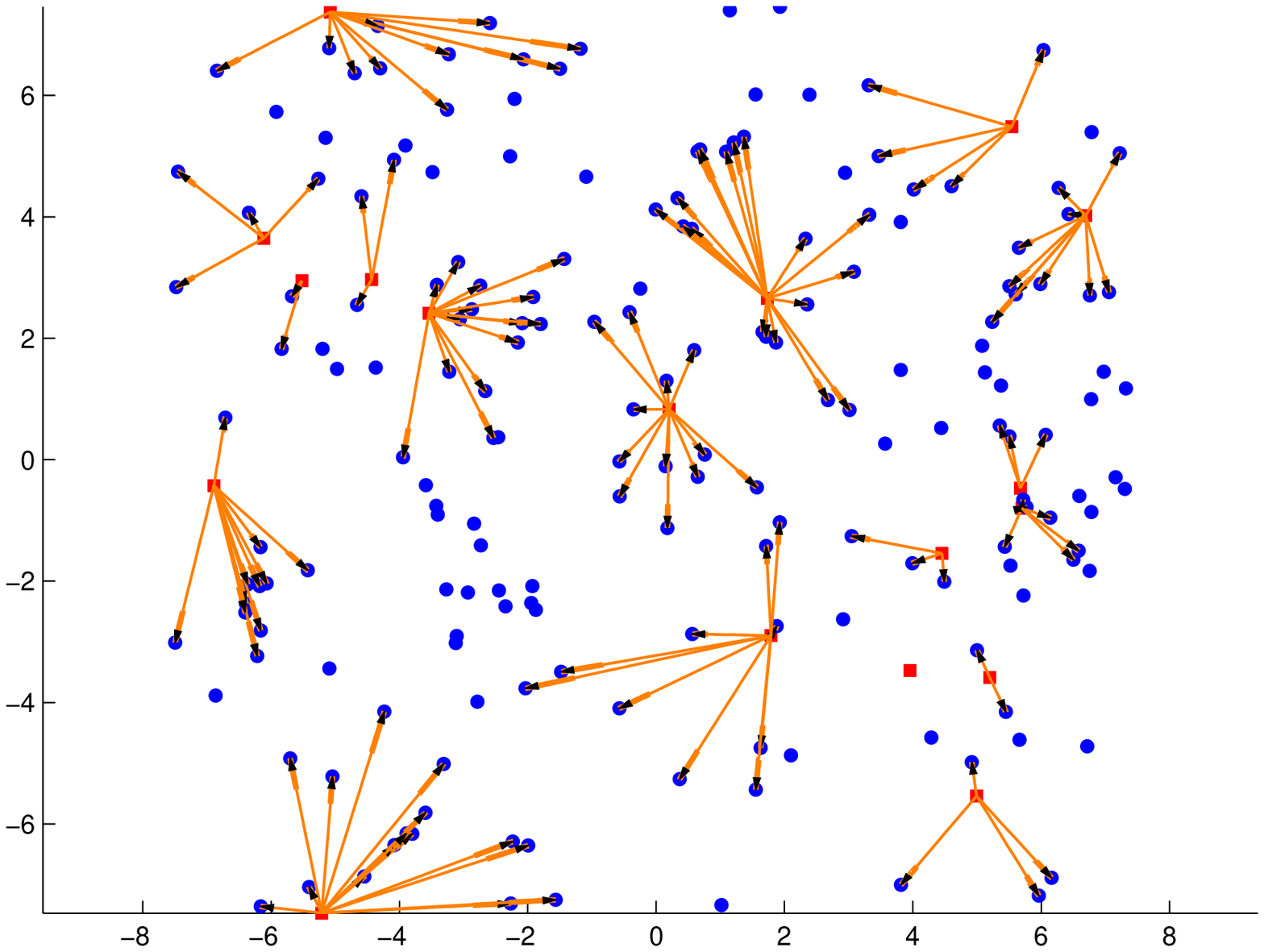}\includegraphics[scale=0.4]{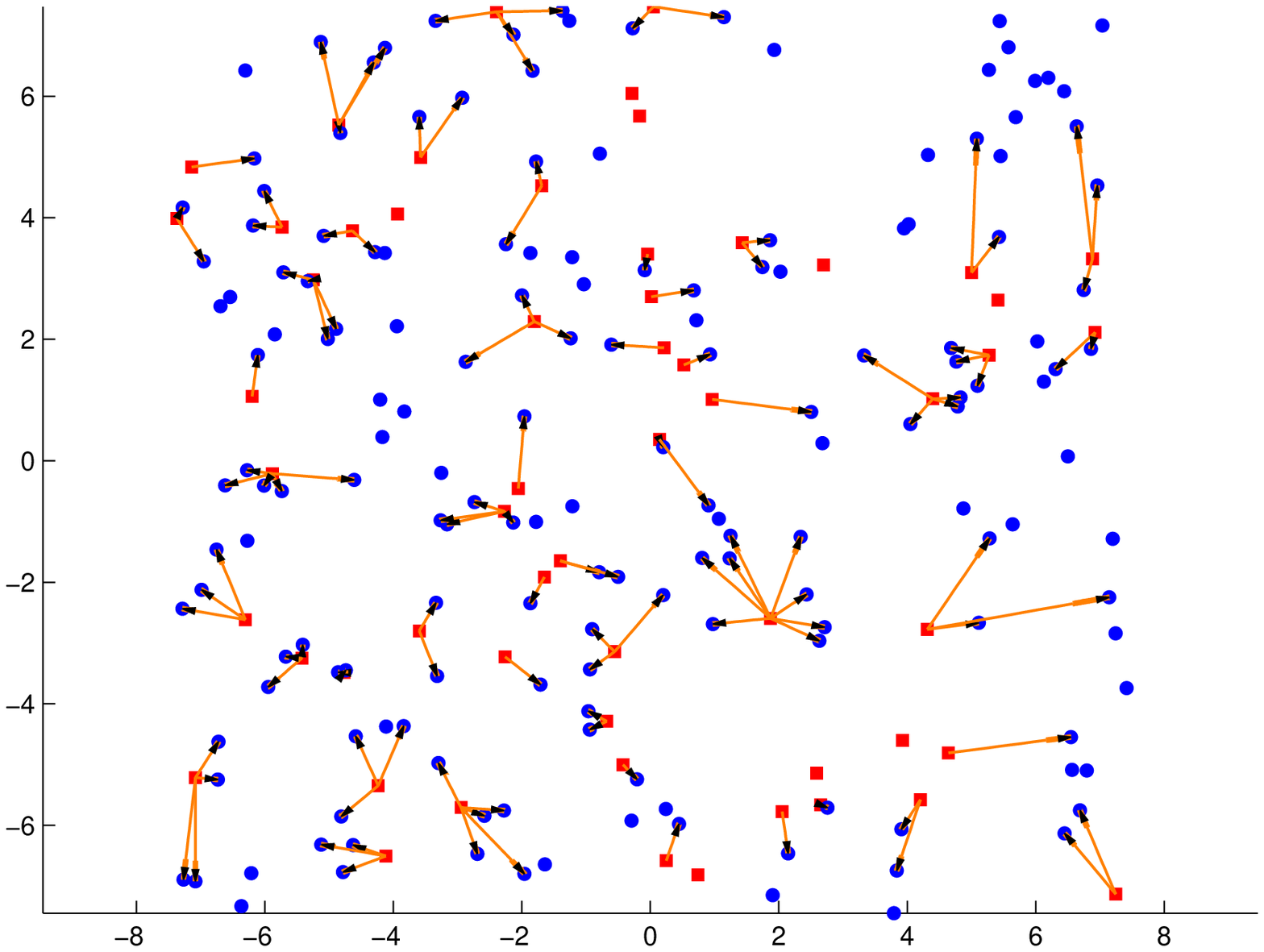} & \tabularnewline
\hline
\end{tabular}
\par\end{centering}
\caption{Illustration of a snapshot graph  $g$ for $p=0.2$ (left) and $p=0.3$ (right) for different realizations of $\phi$. The squares represent
the transmitters and the circles the receivers.}
\label{fig:illus1}
\end{figure}
 See Figure \ref{fig:illus1} for illustration of $g(0)$ and $g(1)$.  Each edge  in this graph $g(k)$   is associated with a  weight  $k$ that represents the time slot in which the edge  was formed. Let
$G(m,n)$ denote the weighted directed multigraph (multiple edges
with different time stamps are allowed between two vertices) formed between
times $m$ and $n>m$, i.e., \[
G(m,n)=\left(\phi,\ \bigcup_{k=n}^{m}E_{k}\right).\]
So $G(m,n)$ is the \emph{edge-union} of the graphs $g(k),$ $m\leq k\leq n$. See Figure  \ref{fig:illus_dynamic}.
\begin{definitn}
A directed path   $x_0,e_0,x_1,$ $e_1,\hdots,e_{q-1},x_{q}$ between the nodes
$x_0, x_q\in \phi $     where $e_i=(x_i,x_{i+1})$ denotes an  edge in the multigraph is said to be a \emph{causal path}
if the weights of the edges $e_i$  are  \emph{strictly increasing} with $i$.%
\end{definitn}
This means that the edge $e_{i-1}$  was formed before $e_{i}$ for $0< i< q$.   For the rest of the paper, we always mean causal path when speaking  about a path.
We observe that the random graph $g(k)$ is a   snapshot of the
ALOHA network at time instant $k$.
The random graph process $G(0,m)$  captures the entire connectivity history up to time $m$.
In the graph $G(0,m)$ there is a notion of time and  causality,
i.e., packets can propagate only on a causal path.

%
%

\section{\label{sec:snapshot}Properties of the   snapshot graph  $g(k)$}

In this section, we will analyze the properties of the random graph
$g(k)$. We first observe that the graphs $g(k)$ are identically
distributed for all $k$. So for this section we will drop the time index  unless  otherwise indicated. %
 $g$ a planar Euclidean graph \emph{even with straight lines} as
edges \cite[Lemma 2]{ganti-2008}. In Figure \ref{fig:illus1},
 realizations of $g$ are shown for $p=0.2$ and $p=0.3$. We first
characterize the distribution of the in-degree of a receiver node
and the out-degree of a transmit node.

\subsection{Node degree distributions}

Let $N_{t}(x)$ denote the number of receivers a transmitter located
at $x$ can connect to, i.e., the out-degree of a transmitting node.
Similarly, let $N_{r}(x)$ denote the number of transmitters that can
connect to a receiver at $x$, i.e., the in-degree of a receiving  node.
We first calculate the average out-degree of a transmitting node.

\begin{prop}
$\mathbb{E}\left[N_{t}(x)\right]=\frac{1-p}{p}\beta^{-2}\left(1-\exp\left(-\lambda
p\pi\beta^{2}\eta^{2}\right)\right) $.
\end{prop}
\begin{proof}
By stationarity of   $\phi$, we  have $N_t(x)\stackrel{d}{=}N_t(o)$ where $\stackrel{d}{=}$ stands for equality in distribution. So it is sufficient to consider the out-degree of a transmitter placed at the origin, which
is given by $\sum_{x\in\phi_{r}}\mathbf{1}(o\rightarrow x,\phi_{t},\r)$.
So the average degree is   \begin{eqnarray*}
\mathbb{E}[N_{t}(o)] & = & \mathbb{E}\left[\sum_{x\in\phi_{r}}\mathbf{1}(o\rightarrow x,\phi_{t},\r)\right]\\
 & \stackrel{(a)}{=} & \lambda(1-p)\int_{\mathbb{R}^2}\mathbb{E}_{\phi_{t}}\left[\mathbf{1}(o\rightarrow x,\phi_{t},\r)\right]\d x\\
 & \stackrel{(b)}{=} & \lambda(1-p)\int_{B(o,\r)}\exp\left(-\lambda p\pi\beta^{2}\|x\|^{2}\right)\d x\\
 & = & \frac{1-p}{p}\beta^{-2}\left(1-\exp(-\lambda p\pi\beta^{2}\r^2)\right),\end{eqnarray*}
where $(a)$ follows from   Campbell's theorem \cite{stoyan} and
the independence of $\phi_{r}$ and $\phi_{t}$. $(b)$ follows from
the fact that $\mathbf{1}(o\rightarrow x,\phi_{t})$ is equal to one
if and only if the ball $B(x,\beta\|x\|)$ does not contain any interferers.
\end{proof}
The average out-degree in the  interference-limited case is obtained by
$\lim_{\r \rightarrow \infty} \mathbb{E}\left[N_{t}(x)\right]$ and is
$\frac{1-p}{p}\beta^{-2}$. Similarly the average out-degree in the noise-limited case is obtained as $\lim_{\beta\rightarrow 0} \mathbb{E}\left[N_{t}(x)\right]$ and is equal to $\lambda(1-p)\pi \r^2$.
\begin{prop}
The probability distribution of $N_{t}$ is given by \begin{equation}
\mathbb{P}\left(N_{t}=m\right)=\sum_{k=m}^{\infty}\frac{(-1)^{k+m}}{k!}\left(\frac{1-p}{p}\right)^{k}V_{k} ,\label{eq:distribution}\end{equation}
where $V_{k} =\int_{B(o,\sqrt{\lambda p}\r)}\cdots\int_{B(o,\sqrt{\lambda p}\r)}\exp\left(-\text{vol}\left(\cup_{i=1}^{k}B(x_{i},\beta\Vert x_{i}\Vert)\right)\right)\d x_{1}\cdots \d x_{k}$.
\end{prop}
\begin{proof}
We provide the complete characterization of $N_{t}$ using the Laplace
transform, given by \begin{eqnarray}
\mathcal{L}_{N_{t}}\left(s\right) & = & \mathbb{E}\left[\exp\left(-sN_{t}\right)\right]\nonumber \\
 & = & \mathbb{E}\left[\exp\left(-s\sum_{x\in\phi_{r}}\mathbf{1}(o\rightarrow x , \phi_t,\r)\right)\right]\nonumber \\
 & \stackrel{(a)}{=} & \mathbb{E}_{\phi_{t}}\exp\left[-\lambda(1-p)\int_{\mathbb{R}^{2}}1-\exp(-s\mathbf{1}(o\rightarrow x,  \phi_t,\r))\d x\right]\nonumber \\
 & = & \mathbb{E}_{\phi_{t}}\exp\left[-\lambda(1-p)(1-\exp(-s))\int_{\mathbb{R}^{2}}\mathbf{1}(o\rightarrow x, \phi_t,\r)\d x\right]\\
&=&\mathbb{E}_{\phi_{t}}\exp\left[-\lambda(1-p)(1-\exp(-s))\int_{B(o,\r)}\mathbf{1}(o\rightarrow x, \phi_t,\infty)\d x\right],\label{eq:equality}\end{eqnarray}
where $(a)$ follows from the probability generating functional of
a PPP.  Let $\nu$ denote a two dimensional Poisson point process of density $1$. {  We then have}
\[\mathbf{1}(o\rightarrow x,\phi_t,\infty) \stackrel{d}{=} \mathbf{1}(o\rightarrow x\sqrt{\lambda p},\nu,\infty).\] Hence
\begin{eqnarray}
\mathcal{L}_{N_{t}}\left(s\right) & = & \mathbb{E}_{\nu}\exp\left[-\frac{ 1-p }{p}(1-\exp(-s))\int_{B(o,\sqrt{\lambda p}\r)}\mathbf{1}(o\rightarrow x,\nu,\infty)\d x\right]\label{eq:lap1}.\end{eqnarray}
 Let $a=\frac{1-p}{p}(1-\exp(-s))$. Then
\begin{eqnarray*}
\mathcal{L}_{N_{t}}\left(s\right) & = & \sum_{k=0}^{\infty}\frac{\left(-a\right)^{k}}{k!}\mathbb{E}_{\nu}\left(\int_{B(o,\sqrt{\lambda p}\r)}\mathbf{1}(o\rightarrow x,\nu)\d x\right)^{k}\nonumber \\
&=&\sum_{k=0}^{\infty}\frac{\left(-a\right)^{k}}{k!}\int_{B(o,\sqrt{\lambda p}\r}\cdots\int_{B(o,\sqrt{\lambda p}\r)}\mathbb{E}_{\nu}\left( \mathbf{1}(o\rightarrow x_1,\nu)\hdots \mathbf{1}(o\rightarrow x_k,\nu) \right)\d x_1\hdots \d x_k\nonumber \\
 & = & 1+\sum_{k=1}^{\infty}\frac{\left(-a\right)^{k}}{k!}\int_{B(o,\sqrt{\lambda p}\r)}\cdots\int_{B(o,\sqrt{\lambda p}\r )}\exp\left(-\text{vol}\left(\cup_{i=1}^{k}B(x_{i},\beta\Vert x_{i}\Vert)\right)\right)\d x_{1}\cdots \d x_{k}\label{eq:prob_nt}
\end{eqnarray*}
 By comparison of coefficients (replace $e^{-s}$ with $z$), we obtain
\eqref{eq:distribution}.
\end{proof}
A lower bound on  $\mathcal{L}_{N_{t}}\left(s\right)$ from \eqref{eq:lap1} is obtained
by using Jensen's inequality: \begin{eqnarray*}
 \mathcal{L}_{N_{t}}\left(s\right) & \stackrel{(a)}{\geq} & \exp\left[-\frac{1-p}{p}(1- e^{-s})\int_{B(o,\sqrt{\lambda p}\r)}\mathbb{E}_\nu\mathbf{1}(o\rightarrow x,\nu,\infty)\d x\right] \\
&\stackrel{(b)}{=}& \exp\left[-\frac{1-p}{p\beta^{2}}(1-e^{-s})(1-e^{-\pi\beta^2 \lambda p \r^2})\right]
 \end{eqnarray*}
where $(a)$  follows from      Jensen's inequality and $(b)$ follows since
$\mathbb{E}_\nu\mathbf{1}(o\rightarrow x,\nu,\infty)$  $=\exp(- \beta^2
\pi \|x\|^2)$.
This is the Laplace transform of a Poisson random variable with mean
$\frac{1-p}{p\beta^{2}}(1-e^{-\pi\beta^2 \lambda p \r^2})$,  which implies  the following lower bound on the probability
of a transmit node being isolated: \begin{eqnarray*}
\mathbb{P}(N_{t}=0) & \geq & \exp\left(-\frac{1-p}{p\beta^{2}}(1-e^{-\pi\beta^2 \lambda p \r^2})\right).\end{eqnarray*}
 We next evaluate the in-degree distribution of a receiving node. Since the point process is stationary, the distribution of $N_r(x)$  is the same for all receivers $x$.
\begin{prop}
The average in-degree $\mathbb{E}[N_{r}(x)]$ of a node in $g$ is $\beta^{-2}(1-e^{-\pi\beta^2 \lambda p \r^2})$.
When $\beta>1$, $N_{r}$ is distributed as a Bernoulli random variable
with mean $\beta^{-2}(1-e^{-\pi\beta^2 \lambda p \r^2})$.
\end{prop}
\begin{proof}
We have $N_r(x)\stackrel{d}{=}N_{r}(o)$ and hence,
\begin{eqnarray*}
\mathbb{E}[N_{r}(o)] & = & \mathbb{E}\left[\sum_{y\in\phi}\mathbf{1}_{\phi_{t}}(y)\mathbf{1}(y\rightarrow o,\phi_t,\r)\right]\\
 & = & \lambda p\int_{\mathbb{R}^{2}}\mathbb{E}_{\phi_{t}}\left[\mathbf{1}(y\rightarrow o,\phi_t,\r)\right]\d y\\
 & = & \lambda p\int_{B(o,\r)}\exp\left(-\lambda p\pi\beta^{2}\|y\|\right)\d y\\
 & = & \beta^{-2}(1-e^{-\pi\beta^2 \lambda p \r^2}).\end{eqnarray*}
 If $\beta>1$,  at most    one transmitter can connect to any receiver, so $N_{r}$ is    Bernoulli.
  Since $\mathbb{E}\left[N_{r}(x)\right]=\beta^{-2}(1-e^{-\pi\beta^2 \lambda p \r^2})$,
we have $N_{r}(x)\sim\text{Bernoulli}(\beta^{-2}(1-e^{-\pi\beta^2 \lambda p \r^2}))$.
\end{proof}
Observe that $\mathbb{E}[N_{t}(x)]$ and $\mathbb{E}[N_{r}(x)]$
are spatial averages and not time averages. We observe that
\[p\mathbb{E}[N_t(o)]= (1-p)\mathbb{E}[N_r(o)],\]
i.e., the time averages of the in-degree and the out-degree are equal.

\subsection{Average time for single-hop  connectivity}
A node may require multiple attempts (time slots) before it is able to
connect to any other node. In this subsection we will consider the time  it
takes for a node to (opportunistically) connect to some other node.
We add a virtual node at the origin and define the number of time slots required
to connect to any node,
        \[ T_O = \min_{k} \left[ \mathbf{1}(o \in \phi_t(k))\prod_{x \in
\phi_r(k)} 1-\mathbf{1}(o\rightarrow x,\phi_t(k),\eta) \right]. \]

\begin{lemma}
  \label{lemma1}
The average single-hop  connection time in a  Poisson network is infinite:
\[\mathbb{E}T_O =\infty.\]
\end{lemma}
\begin{proof}
In the point process $\phi$ the probability that the ball $B(o,\r)$ is empty is equal to $\exp(-\lambda\pi\r^2)$. Hence  a typical transmitter at the origin cannot connect to  any node
 with probability $\exp(-\lambda\pi\r^2)$ regardless of the number of attempts. Hence  $\mathbb{E}T_O =\infty$.
\end{proof}
From the above lemma we observe that the presence of noise which implies a
finite connectivity radius makes the average single-hop connectivity time
infinite. In a Poisson network this happens because the nearest-neighbor
distance is Rayleigh \cite{stoyan} and there exists a positive fraction of nodes with
large nearest-neighbor distance. We now consider an interference-limited
network, i.e., neglect the  finite connectivity radius assumption. Let
$\tilde{T}_O$ denote the opportunistic connectivity time with  the interference
limited assumption.
Let $\tilde{T}_N$ denote the time required for a connection to form between the origin and  its nearest neighbor.  We then have
\[    \tilde{T}_O\leq \tilde{T}_N.\]
\begin{lemma}
\label{lemma_local}
The average time for nearest neighbor connectivity   is   equal to
\[\mathbb{E}\tilde{T}_N = \left\{ \begin{array}{ll}
 (p(1-p) - p^2\nu(\beta))^{-1},& p<\frac{1}{1+\nu(\beta)}\\
\infty, & \text{otherwise}.
\end{array}\right.\]
where \[\nu(\beta)=\left\{ \begin{array}{ll}\beta^2-\pi^{-1}\left\{\beta^2 \cos^{-1}{\frac{\beta}{2}}+\cos^{-1}\left({1-\frac{\beta^2}{2}}\right)-\frac{\beta}{2}\sqrt{4-\beta^2} \right\} &, \beta <2\\
  \beta^2-1&,\beta>2.	\end{array} \right. \]
  \end{lemma}
\begin{proof}
Let $z$ denote the nearest neighbor of  the origin $o$. We first condition on the fact that the node at the origin always transmits and the  node at $z$ always listens.
We then have,
 \begin{eqnarray*}
 \mathbf{1}(o\rightarrow z,\phi_t(k)) &=&  \left[\prod_{x \in \phi\cap
B(o,\|z\|)^c}1-\mathbf{1}(x\in B(z,\beta\|z\|))\mathbf{1}(x\in \phi_t(k))\right]
   \end{eqnarray*}
 The probability that $\tilde{T}_N>k$ is equal to
 \begin{eqnarray}
 \mathbb{P}(\tilde{T}_N>k) &=& \mathbb{E}\prod_{k=1}^k 1- \mathbf{1}(o\rightarrow z,\phi_t(k)).
 \end{eqnarray}

 Let $N(o)$ denote the nearest neighbor of the origin $o$. Conditioning on the point process we have,
  \begin{eqnarray}
 \mathbb{P}(\tilde{T}_N>k\mid \phi, N(o)=z) &=& \left[ 1- \prod_{x \in \phi\cap
B(o,\|z\|)^c}1-\mathbf{1}(x\in B(z,\beta\|z\|))p \right]^k.
  \end{eqnarray}
  So we have
  \begin{eqnarray}
  \mathbb{E}[\tilde{T}_N \mid N(o)=z ]  &=& \mathbb{E}\sum_{k=0}^\infty
\mathbb{P}(\tilde{T}_N>k\mid \phi)\nonumber\\
  &=& \mathbb{E}\left[ \prod_{x \in \phi\cap B(o,\|z\|)^c}1-\mathbf{1}(x\in
B(z,\beta\|z\|))p  \right]^{-1} \nonumber\\
  &=&\exp\left(-\lambda \int_{ B(o,\|z\|)^c} 1- \frac{1}{1-\mathbf{1}(x\in
B(z,\beta\|z\|))p} \d x \right)\nonumber\\
  &=&\exp\left(    \frac{p}{1-p}\lambda \pi   \|z\|^2  \nu(\beta)  \right).
  \label{eq:avg11}
  \end{eqnarray}
  Averaging with respect to the nearest-neighbor  distribution we have
  \begin{eqnarray}
  \mathbb{E}\tilde{T}_N  &=& 2\pi\lambda\int_0^\infty z \exp(-\lambda \pi z^2 ) \exp\left(    \frac{p}{1-p}\lambda \pi   z^2  \nu(\beta)  \right)\d z\\
   &=&\frac{1}{1-p(1-p)^{-1}\nu(\beta)}, \quad  p<\frac{1}{1+\nu(\beta)}.
  \end{eqnarray}
  Removing the conditioning on the node at $o$ transmitting and the nearest neighbor listening,  the result follows.
  \end{proof}
  From the above lemma we observe that there exists a cutoff value for the
ALOHA contention parameter above which $\mathbb{E}\tilde{T}_O=\infty$. See
Figure \ref{fig:cuttoff}. We also observe that the minimum value of
$\mathbb{E}\tilde{T}_N$ occurs at $p=0.5/(1+\nu(\beta))$ and is equal to
$4(1+\nu(\beta))$.
\begin{figure}
\begin{centering}
\includegraphics[width=3.5in]{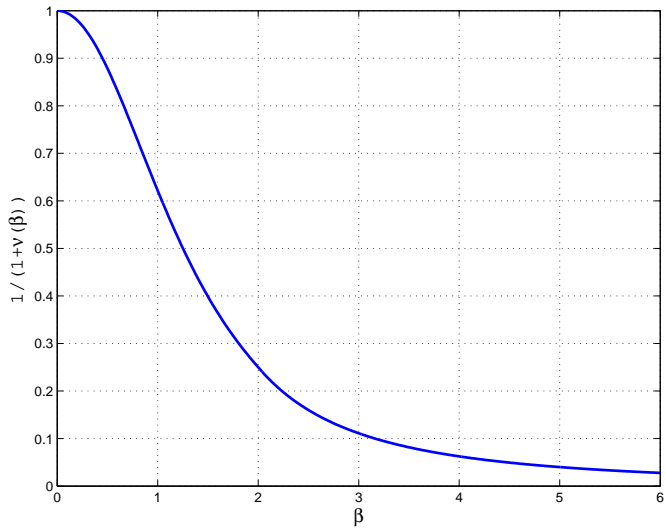}
\par\end{centering}
\caption{The ALOHA parameter $p$ above which the average time for nearest-neighbor connectivity $\mathbb{E}\tilde{T}_N$ is infinite as a function of
$\beta$. }
\label{fig:cuttoff}
\end{figure}

 We now provide a lower bound to the average time required for opportunistic
communication for $\beta>1$.
 \begin{lemma}
   The average time for opportunistic  communication  is  lower bounded by:\\
   $1<\beta <2:$\\
\[\mathbb{E}\tilde{T}_O  > \frac{(\beta-1)^{2}[2+p+(\beta-1)^{2}]}{p(1-p^2)}.\]
$  \beta>2:$\\
\[\mathbb{E}\tilde{T}_O > \left\{ \begin{array}{ll}
 (p - p^2(\beta-1)^2)^{-1},& p<(\beta-1)^{-2}\\
\infty, & \text{otherwise}.
\end{array}\right.\]

 \end{lemma}
 \begin{proof}
 We observe that
 \[\mathbf{1}(o\rightarrow x,\phi_t(k)) \leq  \mathbf{1}(\phi_t(k)\cap B(o,(\beta-1)\|x\|) = \{o\}). \]
 So   the  opportunistic success probability is  upper bounded as
 \begin{equation}
   1- \prod_{x\in \phi_r(k)}1-\mathbf{1}(o\rightarrow x,\phi_t(k)) \leq 1-
   \prod_{x\in \phi_r(k)}1-\mathbf{1}(\phi_t(k)\cap
   B(o,(\beta-1)\|x\|)=\{o\}).\label{eq:upper_time}\end{equation}
 Case 1: $1<\beta <2$.\\
Let $z \in \phi_r$ be the nearest receiver to the origin. We then have
\[B(o,(\beta_2-1)\|z\|) \subset B(o,(\beta_2-1)\|x\|)\quad \forall x\in \phi_r\setminus \{z\}.\]
Hence the success probability at time instant $k$ is bounded by
\[\mathbb{P}(\text{success}\mid \phi ) \leq \mathbb{P}(\phi_t(k)\cap
B(o,(\beta-1)\|z\|)=\{o\}),\]
where $z$ is the nearest node  of $\phi_r(k)$ to the origin.
Let $\eta$  denote the nearest point of the point process $\phi$. Then the right
hand side of the  above equation is equal to the probability that there is
at least one receiver  among the nodes  in the  annulus $A$ centered around the
origin and radius $\eta$ and $\eta/(\beta-1)$. Let  $m$ denote  the number of
nodes of $\phi$ in $A$. We then have
\begin{eqnarray*}
 \mathbb{P}(\phi_t(k)\cap
 B(o,(\beta-1)\|z\|)=\{o\} \mid \phi) &=& 1-p^{m+1}.
\end{eqnarray*}
Hence
\[\mathbb{P}(\tilde{T}_O>n \mid \phi)= p^{(m+1)n}.\]
So we have
\[\mathbb{E}\tilde{T}_O > \mathbb{E}\left[\frac{1}{1-p^{m+1}}\right].\]
Therefore,
\begin{eqnarray}
 \mathbb{E}\tilde{T}_O  &>&  \mathbb{E}\left[\frac{1}{1-p}\mid m=0\right]+\mathbb{E}\left[\frac{1}{1-p^{m+1}}\mid
m>1 \right]\\
&=&\frac{1}{(1-p)(A(\beta)+1)}+\sum_{n=0}^\infty p^n \mathbb{E}[p^{nm}\mid m>0]\\
&=&\frac{1}{(1-p)(A(\beta)+1)}+
2A(\beta)\sum_{n=0}^\infty\frac{p^{2k}}{(A(\beta)+1)(A(\beta)(1-p^k)+1)}\\
&>&\frac{1}{(1-p)(A(\beta)+1)}+\frac{A(\beta)}{(A(\beta)+1)^2(1-p^2)},
\end{eqnarray}
where $A(\beta)= (\beta-1)^{-2}-1$. Multiplying with the average time  for the
origin at $o$ to be a transmitter, we have the result.\\
Case 2: $\beta>2$.
For $\beta>2$, we observe that  the right hand side of \eqref{eq:upper_time}
is equal to $1$ if and only if the  closest point of $\phi$ to the origin $\eta$  is  a
receiver and $B(o,(\beta-1)\eta)$ is devoid of any transmitters.
So we have
\[\mathbb{P}(\text{Success}) < (1-p)^{m+1},\]
where $m$ are the number of points of $\phi$ in the  annulus of radii $\|\eta\|$
and $(1-\beta)\|\eta\|$.  Hence we have
\begin{eqnarray}
  \mathbb{E}\tilde{T}_O  &>&  \mathbb{E}(1-p)^{-m-1}\\
  &=& (1-p)^{-1}\mathbb{E} \exp(\lambda\pi ((\beta-1)^2-1)\eta^2p(1-p)^{-1})\\
  &=&(1-p)^{-1}2\pi\lambda\int_0^\infty x \exp(\lambda\pi
  ((\beta-1)^2-1)x^2p(1-p)^{-1}-\pi\lambda x^2)\d x
\end{eqnarray}
When $ p< (\beta-1)^{-2}$  the last integral converges. Removing the
conditioning on the origin being a transmitter we have the result.
\end{proof}
\begin{figure}
\begin{centering}
\par\end{centering}
\includegraphics[width=3.2in]{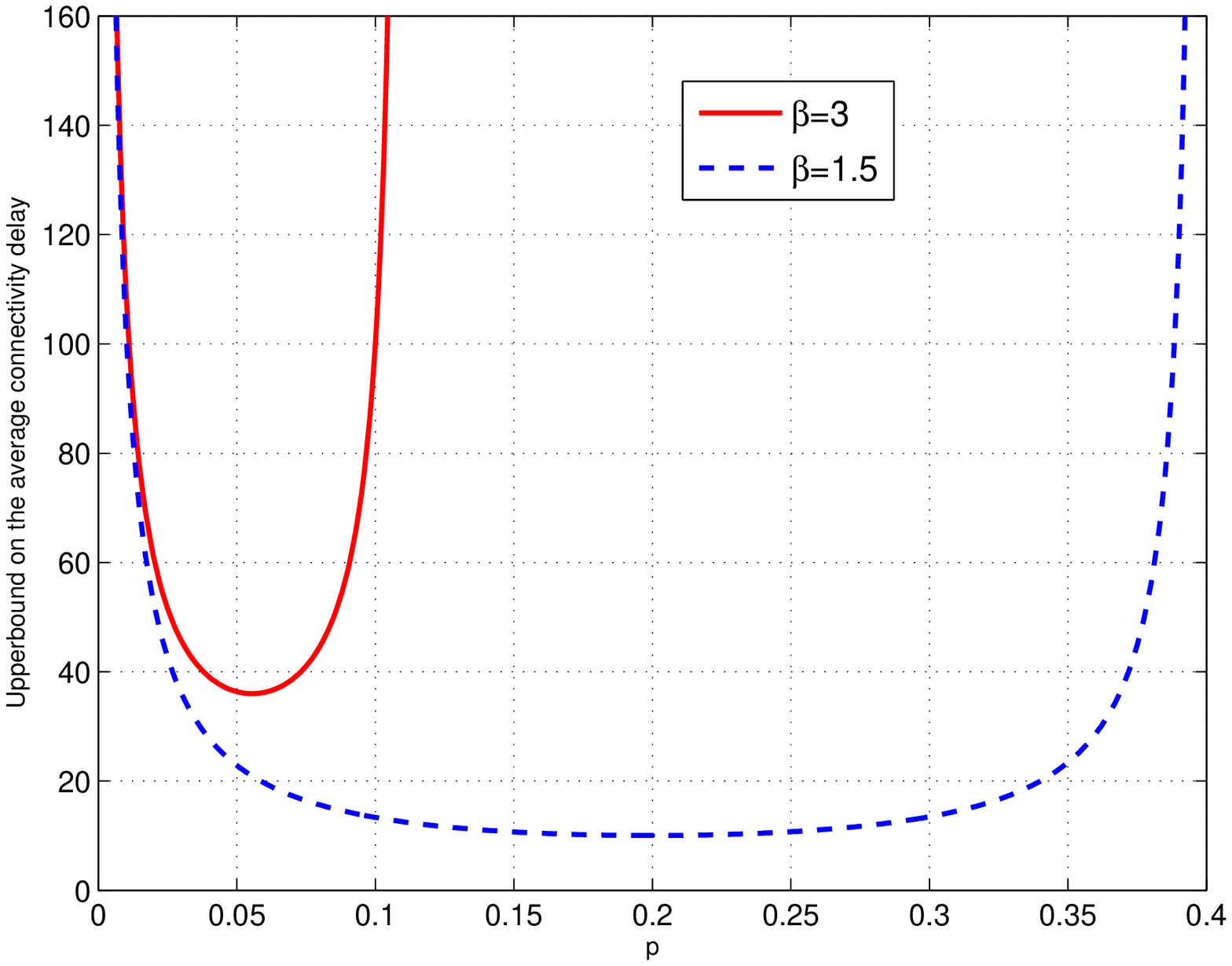}
\includegraphics[width=3.2in]{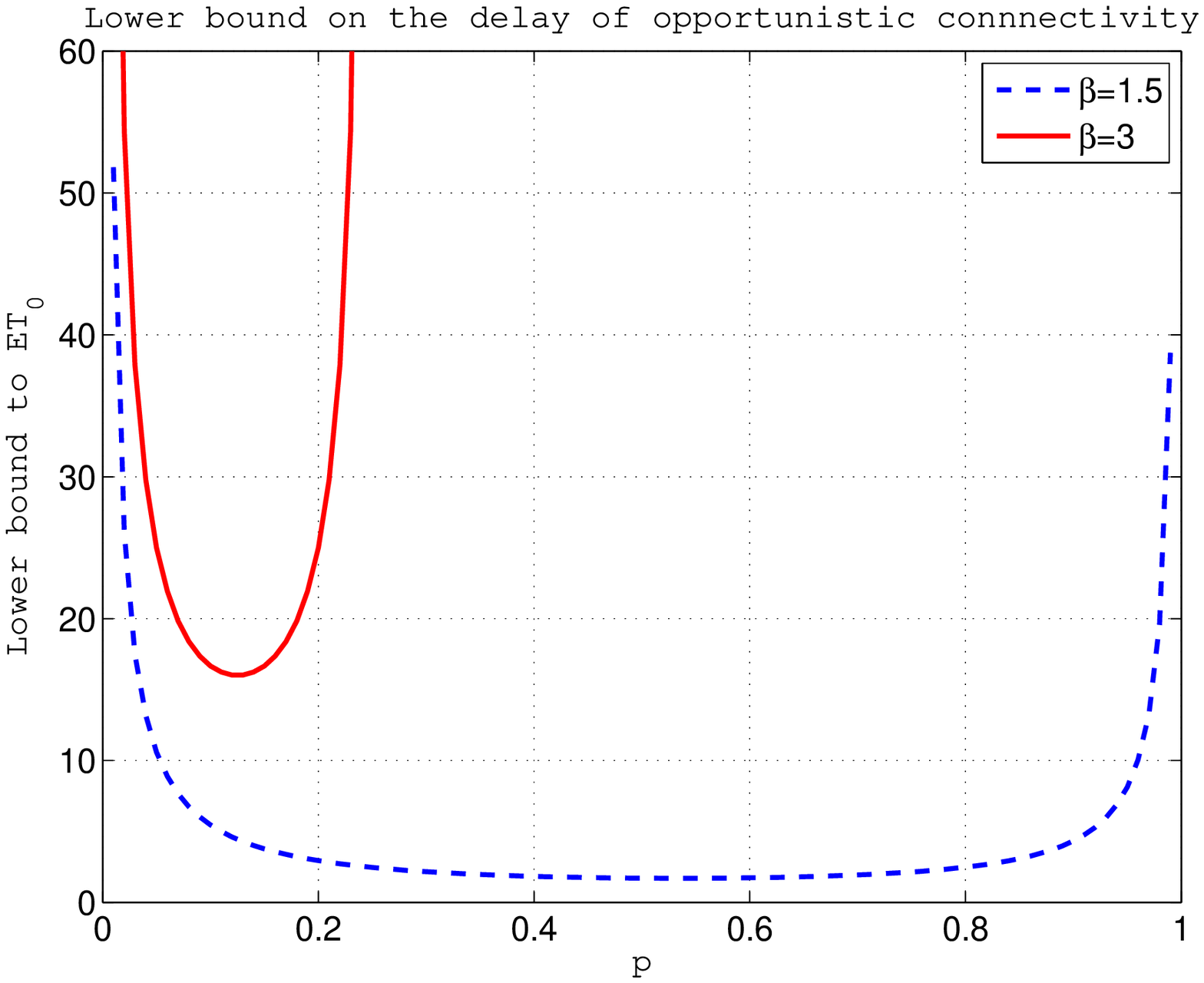}
\caption{ The lower and upper bounds for $\mathbb{E}\tilde{T}_O$ as a function
of $p$ for different values of  $\beta$. The upper bound corresponds to the
average connectivity delay for the nearest-neighbor connectivity
$\mathbb{E}\tilde{T}_N$. }
\label{fig:cuttoff}
\end{figure}

\section{\label{sec:time-evolution} The Time Evolution Graph $G(0,n)$}

In the previous section  we analyzed the snapshot connectivity graph formed
at a particular time instant. In this section we will consider the
superposition of these snapshot graphs and study how the connectivity evolves
over time. %
\subsection{Asymptotic analysis of $G(0,n)$}

We first define the connection time between two nodes. For $x,y\in\phi$,
we denote the \textit{path formation time}  between $x$ and $y$
as \[
T(x,y)=\min\left\{ k\ \colon \ G(0,k)\ \text{has a path from }x\ \text{to }y\right\} .\]
For general $x,y\in\mathbb{R}^{2}$, define $T(x,y)=T(x^{*},y^{*})$
where $x^{*}$ (resp. $y^{*}$) is the point in $\phi$ closest to
$x$ (resp. $y$), with some fixed deterministic rule for breaking
ties (there are no ties almost surely). Since the point process is isotropic, it
is sufficient for most cases to consider destinations   along a given direction.
For notational convenience we define  for  $y\in \mathbb{R}$, $T(x,y)=T(x,(y,0))
$.

This path formation  time is the minimum time required for a packet to propagate from a source $x$
to its destination $y$ in an ALOHA network. In this section we show that
this propagation delay increases linearly with the  source-destination
distance.  Similar to $T(x,y)$ we define \[
T_{n}(x,y)=\min_{k>n}\left\{ k-n\ \colon \ G(n,k)\ \text{has a path from }x\ \text{to }y\right\} .\]
The evolution of the graph $G(0,n)$ is similar to the growth of an epidemic
on the plane, and one can relate the spread of information on the graph $G(0,n)$
to the theory of Markovian contact processes \cite{mollison1978mcp} which was
used to analyze the growth of epidemics. We now provide   bounds on the path
formation time between two points.

In the following arguments we rely on the spatial subadditivity of
$T(o,x)$ to analyze the asymptotic properties. Subadditivity of random
variables is a powerful tool which is often  used to prove results in percolation
and geometric graph theory.
The problem of finding the minimum-delay
path is similar to the problem of   first-passage percolation.
From the definition of $T(o,y)$, we observe that \begin{equation}
T(o,y)\leq T(o,x)+T_{T(o,x)}(x,y).\label{eq:asi:subadd}\end{equation}
We also have that $T_{T(o,n)}(x,y)\stackrel{d}{=}T(x,y)$ from the
way the graph process is defined. Observe that \eqref{eq:asi:subadd}
resembles the triangle inequality (especially if $T_{T(o,y)}(x,y)$
was $T(x,y)$) and thus provides a  pseudo-metric, which holds in FPP problems and  is the reason that the
shortest paths in FPP are called geodesics. In the next two lemmata
we show that the average time for a path to form between two nodes
scales linearly with the distance between them.
\begin{lemma}
\label{lem:The-time-constant}The time constant defined by \[
\mu=\lim_{x\rightarrow\infty}\frac{\mathbb{E}T(o,x)}{x}\]
 exists.
\end{lemma}
\begin{proof}
From \eqref{eq:asi:subadd}, we have \begin{equation}
T(o,y+x)\leq T(o,y)+T_{T(o,y)}(y,y+x).\label{eq:asi:1}\end{equation}
From the definition of the graph, the edge set $E_{k}$ does not depend on
$E_{i},\ i<k$.
Hence $T_{T(o,y)}(y,y+x)$ has the same distribution
as $T(y,y+x)$. Also from the invariance of the point process $\phi$,
we have $T(y,y+x)\stackrel{d}{=}T(o,x)$. Taking expectations of \eqref{eq:asi:1},
we  obtain \[
\mathbb{E}T(o,y+x)\leq \mathbb{E}T(o,y)+\mathbb{E}T(o,x),\]
and the result follows from the basic properties of subadditive
functions. 
\end{proof}
Consistent with the  FPP terminology  we will call
$\mu$ the time constant of the process.
\begin{lemma}
  The time constant for the disc model is infinite,
\[\mu =\infty.\]
\end{lemma}
\begin{proof}
  Follows from Lemma \ref{lemma1}.
\end{proof}
The time constant is infinite because of noise. Because of the finite
connectivity radius a positive fraction of the nodes will not be able to
connect to any other node and hence the time constant is infinite. But
if $\eta >\sqrt{1.435/\lambda} $ \cite{net:Balister05} the disc graph
with radius $\eta$ and node set $\phi$ percolates. Hence there is a giant
connected component
that corresponds to the disc graph formed by just considering the noise  and
not the interference. We denote this giant connected component by   $\Psi_\eta$.

\subsection{Finiteness and positivity of the time constant $\mu$}
We now prove that the any two nodes in this giant component can communicate
in a time that scales linearly with the distance in between. Similar to
$G(0,n)$ we define $G(0,n,\eta)$ as the dynamic graph on $\Psi_\eta$. We can
similarly define for $x,y\in \Psi_\eta$.
\[T(x,y,\eta) =\min\{k:\quad G(0,k,\eta) \text{ has a path from } x \text{ to }
y \}, \] and for $x,y \in \mathbb{R}^2$, $T(x,y,\eta)=T(x^*,y^*,\eta)$ where
$x^*$ and $y^*$ are the points in $\Psi_\eta$ closest to $x$ and $y$. The
following Lemma has been proven in \cite{kong2009connectivity}.
\begin{lemma}
\label{lem:6}
 For $x,y\in\mathbb{R}^2$ and $\|x-y\| <\infty$, $\|x^*-y^*\|<\infty$ almost
surely.
\end{lemma}
We also have the following lemma from \cite{kong2009connectivity} which deals
with the lengths of
the shortest path in terms of the number of hops.
\begin{lemma}
\label{lem:7}
 For $x,y\in \Psi_\eta$, let $L(x,y)$ denote the length (in terms of number of
hops) of the shortest path of the disc graph. If $\|x-y\|<\infty$, then
$L(x,y)<\infty$.
\end{lemma}
We now prove that the time constant is finite and positive on the giant
connected component.
\begin{lemma}
 For any two nodes in $\Psi_\eta$, the average path formation time scales
linearly with the distance, i.e.,
\[0<\mu <\infty,\]
if $0<p<1$.
\end{lemma}
\begin{proof}
 {\em Upper bound:} Let $n$ denote the point $(n,0)$. By subadditivity and
homogenity we have
\[\mathbb{E}T(o,n,\eta)\leq n\mathbb{E}T(o,1,\eta),\]
and hence it is sufficient to show that $\mathbb{E} T(o,1\eta) <\infty$ to
prove $\mu <\infty$. By Lemmata \ref{lem:6} and \ref{lem:7} we have
$L(o^*,1^*)<\infty$ almost surely. Hence the shortest path that connects $0^*$
and $1^*$ in the disc graph has a finite number of edges. Denote the edges by
$e_i, 1\leq i \leq L(o^*,1^*)$ and its corresponding Euclidean length by
$|e_i|$. By the protocol model $|e_i|<\eta$. Let $T_i$ denote the average time for a direct connection to form on
the edge $e_i$. Since the transmitting set of the giant component at time
instant $k$ is a subset of $\phi_t(k)$, the average time obtained in
\eqref{eq:avg11} with $z=\eta$ upper-bounds $T_i$. Hence we have
\[T_i\leq \exp\left(\frac{p}{1-p}\lambda \pi \eta^2\nu(\beta) \right).\]
So
\[\mathbb{E} T(o,1,\eta)< \sum_{i=1}^{L(o^*,1^*)} T_i <
L(o^*,1^*)\exp\left(\frac{p}{1-p}\lambda \pi \eta^2\nu(\beta) \right),\]
which is finite when $p<1$, and hence $\mu <\infty$.

\noindent{\em Lower bound:} By the protocol model any path between $o$ and $n$
should have at least $n/\eta$ hops and hence the average time is always greater
than $n/\eta$ and hence $\mu >0$.
\end{proof}
Hence the information propagation time on the giant component scales linearly
with distance. The fraction of nodes in the giant component increases as the
maximum connectivity distance $\eta$ increases, and hence the set of nodes for
which $\mu<\infty$  increases with increasing $\eta$.

\section{Simulation Results}

\begin{figure}
\begin{centering}
\includegraphics[width=4.2in]{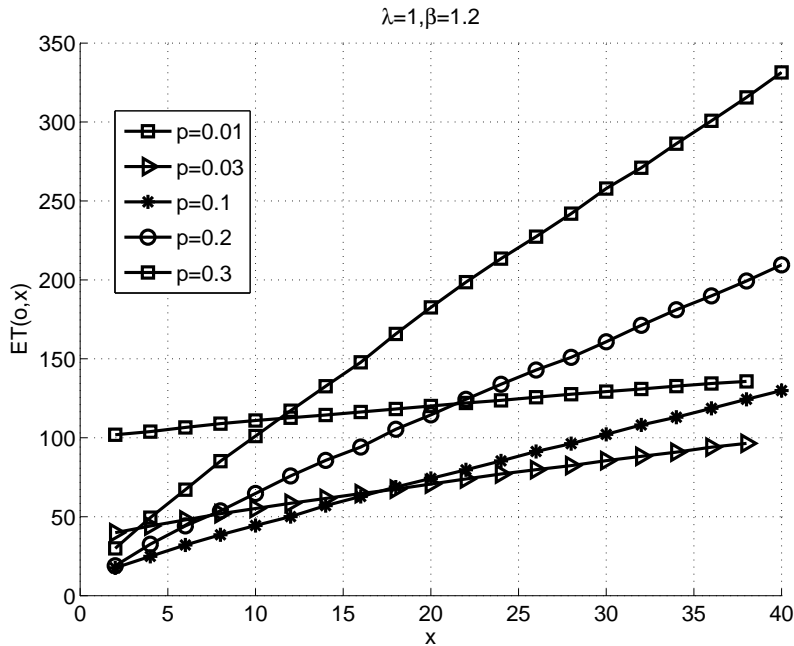}
\par\end{centering}

\caption{$\mathbb{E}T(o,x)$ as a function of $x$, for $\beta=1.2$. We first observe the linear scaling of $\mathbb{E}T(o,x)$ with  the distance $x$ and  that the slope   increases  with $p$. Also for small  values of $x$ we observe that $\mathbb{E}T(o,x) \approx   p^{-1}$  since for small $x$   the path delay time is dominated by the MAC contention time. For small values of $p$, once the source is a transmitter,  long edges form  due to the  low interference.}
\label{fig:avergae_time}
\end{figure}

\begin{figure}
\begin{centering}
\includegraphics[width=4.2in]{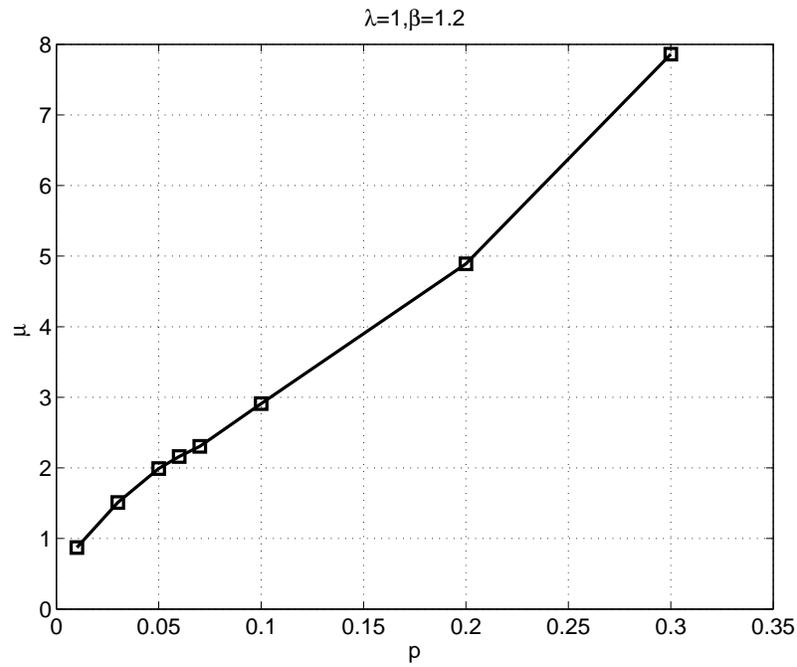}
\par\end{centering}
\caption{The time constant $\mu$ as a function of $p$, for $\beta=1.2$}
\label{fig:time constant}
\end{figure}

\begin{figure}
\begin{centering}
\includegraphics[width=4.2in]{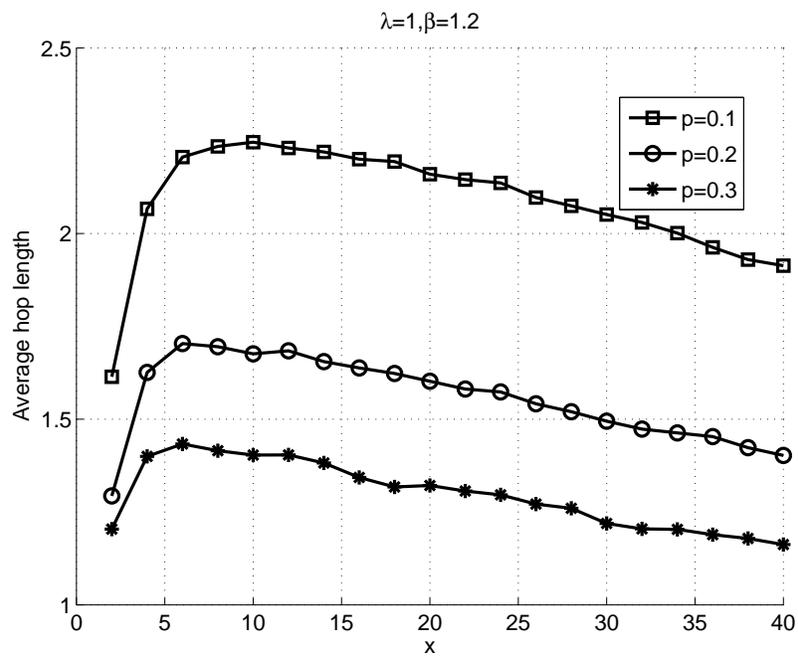}
\par\end{centering}

\caption{Average hop length in the fastest path versus the source-destination distance.}

\label{fig:average_length}
\end{figure}

In this section we illustrate the results using simulation results.
For the purpose of simulation we consider a PPP of unit density in
the square $[-50,50]^{2}$. For most of the simulations, we use $\beta=1.2$, and we average over
 $200$ independent realizations of
the point process.  In Figure \ref{fig:avergae_time},
$\mathbb{E}T(o,x)$ is plotted with respect to $x$ for different
values of $p$. The time constant $\mu$ is plotted as a function
of $p$ in Figure \ref{fig:time constant}. We make the following
observations:
\begin{enumerate}
\item The time constant increases with the ALOHA parameter $p$.
\item In Figure \ref{fig:avergae_time},  we observe that $\mathbb{E}T(o,x)\approx\mu(p)x+C(p)$, where  $C(p)$
is a decreasing function of $p$ and $\mu(p)$   is increasing. For smaller values of $p$,
the time taken for a node to become a transmitter is large, but the
probability of a successful transmission is also high because of the
low density of transmitters. This results in a large $C(p)$ and smaller
$\mu(p)$ for small $p$.
\item Figure \ref{fig:avergae_time} also implies that the presence of interfering
transmitters causes the delay to increase when the packet has to be
transmitted over longer distances. So when the   packet transmission
distance is large, it is beneficial to decrease the density of contending transmitters.
\item For each $x$, there is an optimal $p$ which minimizes the delay, and the optimum $p$ is a decreasing function of $x$.
\end{enumerate}
For two nodes located at $o$ and $x$ and $\|x\|$ large, there  will in general
be many  paths between $o$ and $x$ which form by time $\mu\|x\|$.
 From such an ensemble of delay-optimal paths, we will consider paths which have the minimum
number of hops and call them \textit{fastest paths}.  In Figure
\ref{fig:average_length}, we show the average number of  hops
  in these paths. We observe that for a given $p$,
the average hop length decreases as the source-destination distance
$x$ increases. This shows that for larger source-destination distance,
it is beneficial to use shorter hops since they are more reliable
and form faster than longer hops. Also from Figure \ref{fig:time constant},
we observe that for larger $x$, it is beneficial to be less aggressive in terms of spatial reuse
and use a smaller $p$.

%
%
%
%
%

%
\section{Conclusions}
Connectivity   in a wireless network is dynamic and directed because of
the MAC scheduler  and the half-duplex radios. Since these properties are not
captured in  static graph models that are usually used,  we have introduced  a
dynamic connectivity graph and analyzed its properties for ALOHA.   We have
shown that the time taken for a causal path to form between a source and a
destination on this  dynamic ALOHA graph scales linearly with the
source-destination distance for large fraction of nodes. The fraction of
nodes for which the time-constant is finite increases with increasing power.
So we can state the following: {\em Networks are inherently noise-limited (or
power-limited) as given sufficient time, the MAC protocol can induce enough
randomness to deal with the interference. }
 By simulations we showed that it is
beneficial to use higher value of the ALOHA contention parameter for smaller
source-destination distances and lower value for large distances, and that the
average hop length of the fastest paths   first increases rapidly  but then
decreases slowly as a function of the source-destination  distance. These
observations provide some insight  how to choose the hop length  for efficient
routing in ad hoc networks.
\section*{Acknowledgments}
The partial support of NSF (grants CNS 04-47869, CCF 728763) and the DARPA/IPTO IT-MANET program
(grant W911NF-07-1-0028) is gratefully acknowledged.
\bibliographystyle{ieeetr}
\bibliography{point_process}

\end{document}